\documentclass[preprint]{aastex}
\usepackage{natbib}
\usepackage{amsmath}
\begin{document}

\title{A Model for Radio Emission from Solar Coronal Shocks}

\author{G. Q. Zhao\altaffilmark{1,2}, L. Chen\altaffilmark{1}, and D. J. Wu\altaffilmark{1,3}}
\affil{$^1$Purple Mountain Observatory, CAS, Nanjing 210008, China}
\affil{$^2$University of Chinese Academy of Sciences, Beijing 100049, China}
\altaffiltext{3}{Correspondence should be sent to: djwu@pmo.ac.cn.}

\begin{abstract}
Solar coronal shocks are very common phenomena in the solar atmosphere and are believed to be the drivers of solar type II radio bursts. However, the microphysical nature of these emissions is still an open problem. This paper proposes that electron cyclotron maser (ECM) emission is responsible for the generation of radiations from the coronal shocks. In the present model, an energetic ion beam accelerated by the shock excites first Alfv\'en wave (AW) and then the excited AW leads to the formation of a density-depleted duct along the foreshock boundary of the shock. In this density-depleted duct, the energetic electron beam produced via the shock acceleration can effectively excite radio emission by the ECM instability. Our results show that this model may have potential application to solar type II radio bursts.
\end{abstract}


\section{Introduction}

A shock standing ahead of a magnetic structure ejected from the Sun is a very common phenomenon in solar and interplanetary physics. This idea of the shock was proposed by \citet{gol55p03,gol62p00}, which was confirmed subsequently via in situ observations \citep{son64p53}. In the solar corona, the shock is believed to be driven by fast coronal mass ejection \citep{man02p67,lar03p16,cli04p05,liu09p51,che11p01,ram12p07}, or by the pressure pulse of a flare \citep{har65p61,gop98p07,nio11p31,mag08p05,mag10p66,mag12p52}. The early known signatures of coronal shocks are type II radio bursts which often appear in a form of two intense emission bands drifting gradually from higher to lower frequencies as revealed from the solar radio dynamic spectra \citep{pay47p56,wil50p87,nel85p33}. Other evidences of coronal shocks are suggested in terms of Moreton waves \citep{mor60p94,mor60p57,che02p99}, streamer deflections \citep{gos74p81,mic84p11,she00p81}, and magnetohydrodynamic simulations \citep{che00p75,man04p07}. In recent studies, the existences of coronal shocks have been shown by direct observations both in white light \citep{vou03p92,ont09p67,gop09p27,gop11p17,she13p43} and the extreme ultraviolet \citep{bem10p30,koz11p25,mas11p60,gop12p72}. One can refer to the recent reviews for details of theory and properties of shocks \citep{vrs08p15,tre09p09,war10p27,pat12p87}.

For the coronal shock, one of the most challenging and fascinating questions is how it can result in emission of radiation. Two elements are required for the emission of radiation of the shock. The first one is the energetic electrons produced by the shock, and the second one is the emission mechanism by which part of kinetic energy of the energetic electrons can be converted to radiation. For the energetic electrons, many authors found that their acceleration mechanism, in terms of shock drift acceleration \citep{hol83p37,ste84p93,bal01p61} or fast fermi acceleration \citep{wuc84p57,ler84p49}, is actually simple and effective for the case of nearly perpendicular shocks (NPSs). The key point is that these NPSs can act as fast-moving magnetic mirrors which reflect upstream electrons along the ambient magnetic field, and consequently one can obtain energetic electrons characterized by beam with a loss-cone \citep[or "hollow beam", "ring beam" in literatures;][]{wuc86p92,man05p19,yoo07p01}. For the emission mechanism two theories, related to induced (or coherent in literatures) emissions, have been proposed. One is so called plasma-emission suggested by \citet{gin58p53}. This theory posits that electrostatic waves are first generated by instabilities and then are converted to electromagnetic waves via nonlinear wave$-$wave interaction \citep{mer85p77,ben88p67}. Here, both the high level of electrostatic waves and conversion efficiency are required to explain the unusually high brightness temperature (up to $10^{13}$ K) from the observations of type II bursts \citep[][and references therein]{nel85p33}. The other is the theory of direct excitation of radiations in which electromagnetic waves can be amplified efficiently by perpendicular free energy of energetic electrons via wave$-$particle interaction \citep{twi58p64,wuc85p15}. Based on this theory, for nearly perpendicular coronal shocks two papers have been presented to show the emission processes related to synchrotron maser emission \citep{wuc86p92} or electron cyclotron maser emission \citep[ECME;][]{yoo07p01}. ECME from interplanetary shocks or other quasi-perpendicular astrophysical shocks were also discussed \citep[see,][]{far01p01,bin03p79}.

ECME is a well-known emission mechanism and has been extensively discussed as a dominant mechanism of producing high-power radiation in magnetized plasmas \citep[e.g., see a review by][]{tre06p29}. In the discussion of ECME a low-density region, with plasma frequency lower than electron cyclotron frequency, is important so that the energetic electrons with perpendicular free energy can efficiently drive ECME \citep{bin03p79,lee13}. This low-density region, if the ECME is responsible for the radio emission of coronal shocks, should be necessary for the ECME to effectively emit radiations near both the fundamental (F) and its second harmonic (H). This is because observations showed that two emission bands (F and H) of type II bursts in general have their frequency ratio of about $1:2$ \citep[e.g.,][]{nel85p33,man95p75}.

However, the existence of a low-density region related to coronal shocks, to the best of our knowledge, has not been discussed. Only a relevant study of the depletion of plasma density in a flux tube in the solar corona was presented \citep{wuc06p17}. According to the study by \citet{wuc06p17}, Alfv\'en waves (AWs) excited by ion beam can deplete plasma density and result in the formation of a density-depleted duct on the path of the ion beam traveling due to the pressure of the AWs. This process is expected to be effective in a low-beta plasma in which the magnetic pressure dominates the plasma pressure \citep[e.g.,][]{dul85p19}.

The present paper is devoted to further reveal the physical processes of radio emission from solar coronal shocks based on ECME, in which AWs generated by ion beam are taken into account self-consistently. The paper is organized as follows. In Section 2 the basic physical model is given, in which the ion and electron beams in the foreshock boundary, excitation of AWs by ion beams, density depletion by AWs, and ECME in the presence of AWs are described, respectively. The calculated results based on solar coronal parameters are presented in Section 3. Several characters of the present model related to type II radio bursts are discussed in Section 4. Finally, the conclusions with some brief discussion are given in Section 5.

\section{Basic physical model}

\subsection{Ion and electron beams in the foreshock boundary}

Many authors state that NPSs can accelerate ions as well as electrons \citep[e.g., see a review by][]{tre09p09}. This is conceivable on the basis of extensive studies of the nearly perpendicular Earth's bow shock by both theories and observations. Early observations showed that a thin sheet with energetic ion and/or electron spikes lies just behind the edge of interplanetary magnetic field lines almost parallel to the shock surface \citep[e.g.,][]{sar75p33,and79p401,and81p45}. In fact, ions as well as electrons in NPSs are expected to be subject to shock drift acceleration process \citep[e.g., see a review by][]{bal01p61}, and high energy field-aligned beams at the upstream edge of the foreshock can be obtained \citep[e.g.,][]{fus95p43}. Such a process was further revealed by the in situ observations of the bursts of energetic ions close to the foreshock boundary of the Earth's bow shock \citep{mez99p25,mez02p43} and was demonstrated by numerical simulations \citep{lev01p67,yan09p11}. Hence, we adopt that ion and electron beams can exist near the nearly perpendicular coronal shocks.

\begin{figure}
\epsscale{0.8} \plotone{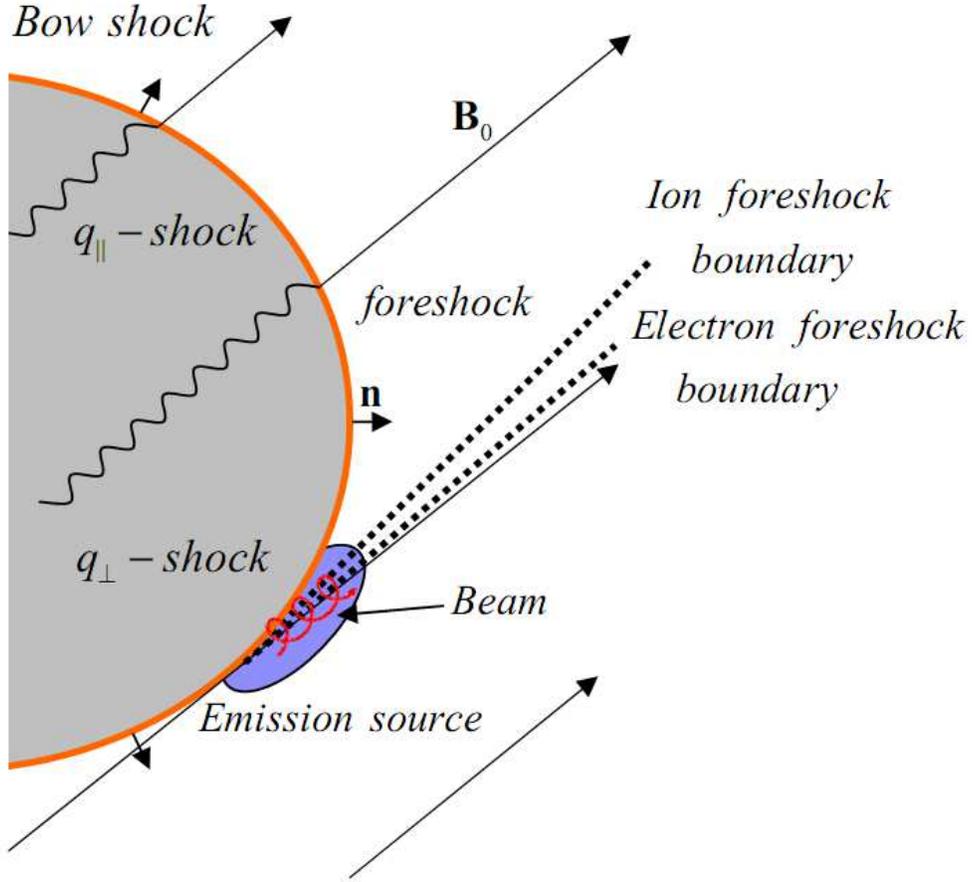} \caption{Schematic describing a propagating bow shock, foreshock and emission source region. The $q_\parallel$-shock and $q_\perp$-shock represent the quasi-parallel shock and the quasi-perpendicular shock, respectively. The ion and electron foreshock boundaries are denoted by two dashed lines. The emission source region, i.e., the vicinity of tangency point, is shown as the blue area where the ion and electron foreshock boundaries are nearly superimposed.\label{fig1}}
\end{figure}

For illustration let us consider a propagating coronal bow shock as described in Figure 1. The shock may consist of quasi-parallel and quasi-perpendicular parts in terms of the shock normal angle, i.e., ${{\theta}_{Bn}}$, which is defined by the angle between the upstream magnetic field (${\mathbf{B}_{0}}$) and the local shock normal vector ($\mathbf{n}$). The shock geometry is called quasi-parallel when ${{\theta}_{Bn}}<{{45}^{\circ}}$ and quasi-perpendicular when ${{\theta}_{Bn}}>{{45}^{\circ}}$. The space from the bow shock to just downstream of the tangent magnetic field lines is known as the foreshock, which is permeated by particles backstreaming from the shock. Both ions and electrons observed close to their foreshock boundaries, denoted by two dashed lines in Figure 1, are usually characterized by beam and have the highest energy \citep{eas05p41}. The main focus of our interest is thus on these ions and electrons close to their foreshock boundaries in this paper. In particular, there is a region where the upstream magnetic field is nearly tangential to the bow shock. We suggest that this region is emission source region, which is marked by blue area where the ion and electron foreshock boundaries are nearly superimposed. In such a region three processes of (1) excitation of AWs by ion beams, (2) depleting density and forming a duct, (3) ECME in the duct can be expected, which will be demonstrated in the following subsections.

\subsection{Excitation of AWs by the ion beam}

A large number of low frequency waves may be generated due to free energy of the ion beam \citep{tsu81p17,rus83p55,gar85p42,gar91p73,bri91p11}. Among these waves, AWs are outstanding, which in general tend to be undamped and long-lived in a plasma \citep[]{bel71p34}. Excitation of AWs by the ion beam can be extensively carried out in terms of beam instability or spontaneous process \citep[][and references therein]{wud12p11}. Here, for the sake of simplicity, we employ the scheme introduced by \citet{has82}. We assume that the ion (i.e., proton) beam is tenuous ($n_{bi} \ll n_0$) and fast ($v_{bi} \gg v_A$) moving along the ambient magnetic field in the $z$-direction, where $n_0$ and $v_A$ are the ambient density and Alv\'en velocity, respectively. For a low-beta plasma, the dispersion equation of AW can be written as \citep{has82}
\begin{eqnarray}
k_{Az}^{2}-\frac{{{w}^{2}}}{v_{A}^{2}}-\frac{{{n}_{bi}}}{{{n}_{0}}}\frac{{{\Omega }_{i}}}{v_{A}^{2}}\left( w-{{k}_{Az}}{{v}_{bi}} \right)=0,
\end{eqnarray}
where $\Omega_{i}$ is the ion gyrofrequency, and in obtaining Equation (1) we have considered AW propagating parallel to the ambient magnetic field with the wave vector $\mathbf{k}_A = (0, 0, k_{Az})$.

The solution of the Equation (1) can be given as follows:
\begin{eqnarray}
\frac{w}{{{k}_{Az}}{{v}_{A}}}={{\alpha }_{b}}\pm \sqrt{1+\alpha _{b}^{2}-2{{\alpha }_{b}}\frac{{{v}_{bi}}}{{{v}_{A}}}},
\end{eqnarray}
with the parameter
\begin{eqnarray}
{{\alpha }_{b}}\equiv -\frac{1}{2}\frac{{{n}_{bi}}}{{{n}_{0}}}\frac{{{\Omega }_{i}}}{{{k}_{Az}}{{v}_{A}}}.
\end{eqnarray}
For the case of $\alpha_b > 0$, i.e., $k_{Az} < 0$ implying that the AW propagates actually in the opposite direction of the ambient magnetic field \citep{sen81p87,gom98p83}, it is clear that, when the condition
\begin{eqnarray}
\frac{{{v}_{bi}}}{{{v}_{A}}}>\frac{1+\alpha _{b}^{2}}{2{{\alpha }_{b}}}\ge 1
\end{eqnarray}
is satisfied, the AW mode becomes unstable and is excited.

\subsection{Density-depleted duct by the excited AWs}

In a low-beta plasma, the excited AW will deplete the density through magnetic compression \citep{wuc06p17}, and a density-depleted duct can be expected along the whole ion foreshock boundary where the beam has the highest velocity. For the discussion we introduce the parameter of density inside the duct, i.e., $n_D$. As the AW grows, the density inside the duct will reduce and the local Alfv\'en velocity $v_{AD}$ increases. When $v_{AD}$ reaches $v_{bi}$, the AW has its maximal level since instability condition of Equation (4) is no longer satisfied. In particular, one can find that the density depletion becomes considerable according to the relation
\begin{eqnarray}
\frac{n_D}{n_0}=\frac{v_A^2}{v_{bi}^2},
\end{eqnarray}
which implies $n_D \ll n_0$ under the condition of $v_{bi} \gg v_A$. Here the maximal level of the AW ($B_w^2$) can be estimated by the pressure balance between the inside and the outside of the duct. that is
\begin{eqnarray}
\frac{B_0^2}{8\pi}+\frac{B_w^2}{8\pi}+n_D{T_D}=\frac{B_0^2}{8\pi}+n_0{T_0}
\end{eqnarray}
where $T_D$ and $T_0$ are the kinetic temperature inside and outside the duct, respectively. The $T_D$ is in general slightly larger than $T_0$ due to the possible heating effect (e.g. the possible wave heating) caused by the ion beam in the duct. The Equation (6) leads to
\begin{eqnarray}
\frac{B_w^2}{B_0^2}=(1-\frac{n_D{T_D}}{n_0{T_0}})\beta_0=(1-\frac{v_A^2{T_D}}{v_{bi}^2{T_0}})\beta_0 \lesssim \beta_0
\end{eqnarray}
where $\beta_0 \equiv 8\pi{n_0}{T_0}/{B_0^2}$ is the ambient plasma beta. Hence one can find that the AW level relative to the square of the ambient magnetic field may approach, but not exceed, the ambient plasma beta in this discussion.

\subsection{ECME in the density-depleted duct}
In the density-depleted duct, the condition of $\omega_{pe}/\Omega_{e} \lesssim 1$ may be fulfilled, which is in favor of ECME, where $\omega_{pe}$ and $\Omega_{e}$ are the electron plasma frequency and gyrofrequency, respectively. Here we will use the new ECME found by \citet{wuc12p02} in which the presence of AWs was considered. AWs can significantly influence the basic physics of wave$-$particle interaction. This is because the velocity of each electron becomes oscillational along a uniform ambient field in the presence of AWs, and the additional longitudinal oscillation modifies the usual resonant condition according to linear kinetic theory. This new ECME was further explored subsequently and the numerical results showed that the growth rate of the ordinary (O) mode wave can greatly increase in the presence of AWs \citep{zha13p31,zha13p75,zha13p03}.

Taking into account the effect of AWs, the formulations of the growth rate of electromagnetic waves are given by \citet{wuc12p02}, They are
\begin{eqnarray}
 {\gamma _k} & = & \frac{{{n_{be}}}}{{{n_0}}}\pi \frac{{\omega _{pe}^2}}{2}\sum\limits_n^{} {\sum\limits_q^{} {} } \int\limits_{}^{} {{d^3}} {\bf{v}}v{\mu ^2}\left( {1 + \varepsilon q} \right)J_n^2({p_{}}){J_q}({\rho _{}}) \nonumber \\
 & & \times \delta \left[ {{\omega _k} - \left( {n + q} \right){\Omega _e}/\gamma  - {k_z}\mu v} \right]\left( {\frac{{\partial {F_{be}}({\bf{v}})}}{{\partial {v_{}}}} - \frac{\mu }{v}\frac{{\partial {F_{be}}({\bf{v}})}}{{\partial \mu }}}
 \right)
 \end{eqnarray}
for O mode and
\begin{eqnarray}
 {\gamma _k} & \approx & \frac{{{n_{be}}}}{{{n_0}}}\pi \frac{{{{\left( {\omega _k^2 - \Omega _e^2} \right)}^2}}}{{2\omega _k^2}}\sum\limits_n^{} {\sum\limits_q^{} {} } \int\limits_{}^{} {{d^3}} {\bf{v}}v{\left( {1 - {\mu ^2}} \right)_{}}{[ J_n^{'}({p_{}}) ]^2}{J_q}({\rho _{}}) \nonumber \\
 & &\times \delta \left[ {{\omega _k} - \left( {n + q} \right){\Omega _e}/\gamma  - {k_z}v\mu } \right]\left( {\frac{{\partial {F_{be}}({\bf{v}})}}{{\partial {v_{}}}} - \frac{\mu }{v}\frac{{\partial {F_{be}}({\bf{v}})}}{{\partial \mu }}}
 \right)
 \end{eqnarray}
for X mode, where $n_0$ and $n_{be}$ are electron number densities of the ambient plasma and nonthermal component, respectively; $p = {k_ \bot }{v_ \bot }/{\Omega _e}$, $\rho =
{k_z}v\mu B_w^2/B_0^2{\Omega _e}$, $\varepsilon  =
{\Omega_e}/{k_z}v\mu$, and $\mu  = {v_ z }/{v}$; $k_\bot$ and $k_z$ are
the components of the emitted wave vector $\bf{k}$ perpendicular and
parallel to the ambient magnetic field; $\omega_k$ is the emitted wave frequency; ${J_n}(p)$ and ${J_q}(\rho)$ are
Bessel functions of order of $n$ and $q$, respectively; $J_n^{'}(p)$
is the derivative; $\gamma  = {\left( {1 - {v^2}/{c^2}} \right)^{
-1/2}}$ is the relativistic factor; ${F_{be}}({\bf{v}})$ is the distribution function of energetic electrons which will be given later; Finally, the Dirac function $\delta \left[ {{\omega _k} - \left( {n + q} \right){\Omega _e}/\gamma - {k_z}v\mu } \right]$ implies the new resonant condition. This new resonant condition is attributed to the electron oscillational motion along the ambient magnetic field in the presence of AWs.

\begin{figure}
\epsscale{0.8} \plotone{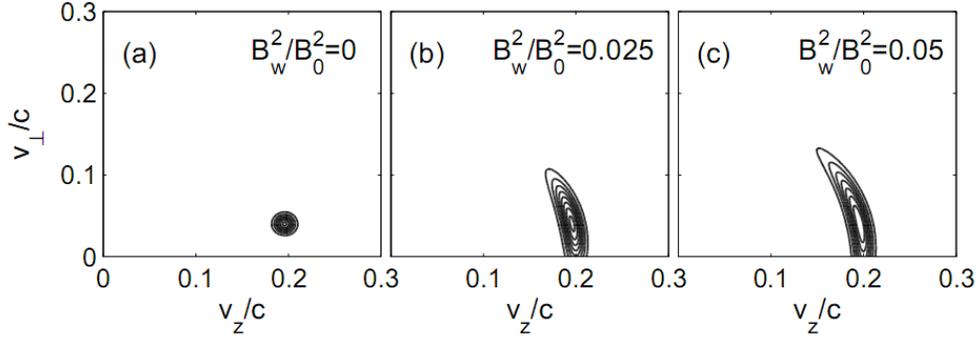} \caption{Contour plot of the beam
distribution function described by the Equation (10) for the cases of $B_w^2/B_0^2=0$, $B_w^2/B_0^2=0.025$, and $B_w^2/B_0^2=0.05$. Panel (a) describes a ring-beam distribution while panels (b) and (c) show the crescent-shaped distribution in the presence of AWs. \label{fig2}}
\end{figure}

\begin{figure}
\epsscale{0.8} \plotone{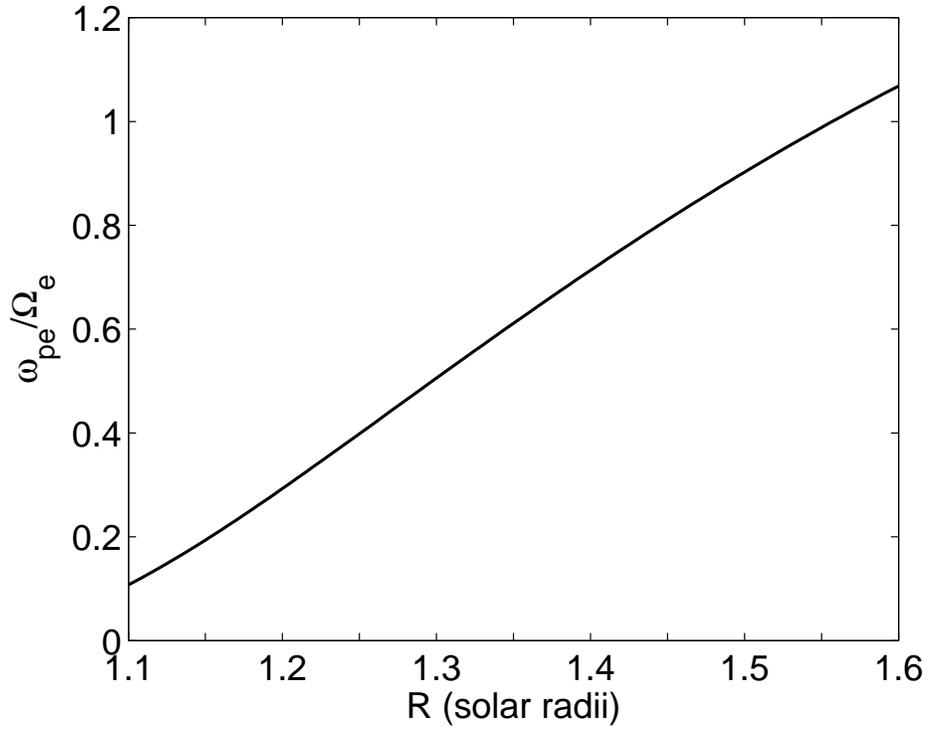} \caption{$\omega_{pe}/\Omega_e$ vs. the heliocentric distance in the duct. \label{fig3}}
\end{figure}

To model the distribution function of energetic electrons, we first consider that the energetic electrons are characterized by a ring-beam. Taking account of the presence of AWs, we also consider that these electrons will be subject to pitch-angle scattering process by small amplitude AWs and a crescent-shaped distribution is formed. This process is first introduced for ions \citep{wuc97p56,wan06p01}, while it also applies to electrons as confirmed by test-particle simulations \citep{luq06p69,wuc12p02}. Hence, the distribution function may be described by
\begin{eqnarray}
{{F}_{be}}(v,\mu )=D\exp \left( -\frac{{{(v-{{v}_{be}})}^{2}}}{{{\alpha }^{2}}}-\frac{(\sqrt{1-{\mu}^{2}}-\nu_0)^2}{{{\Delta }^{2}}} \right),
\end{eqnarray}
where $D$ is the normalized factor; $v_{be}$ is the beam velocity; $\nu_0=v_r/v_{be}$ and $v_r$ denotes the ring velocity; $\alpha$ and $\Delta$ represent the velocity dispersion and the pitch-angle dispersion, respectively; and $\Delta$ is given by
\begin{eqnarray}
\Delta = \frac{\alpha}{v_{be}}\sqrt{1+2\frac{B_w^2}{B_0^2}\frac{v_{be}^2}{\alpha^2}}
\end{eqnarray}
with $\alpha=0.05 v_{be}$ in this paper. It is noted that $\Delta=\alpha/v_{be}$ for the case of no AWs (${B_w}=0$) while $\Delta \simeq \sqrt{2{B_w^2}/{B_0^2}}$ for a finite AWs level \citep{wuc12p02}, since ${v_{be}^2}/{\alpha^2} \gg 1$ is in general fulfilled. To graphically show the electron distribution, we present Figure 2 for three cases of ${B_w^2}/{B_0^2}=0$, ${B_w^2}/{B_0^2}=0.025$ and ${B_w^2}/{B_0^2}=0.05$, where the parameters $v_{be}=0.2c$  ($c$ is the speed of light) and $\nu_0=0.2$ have been used. One can see that the pitch-angle scattering diffuses the velocity distribution into a crescent-shaped configuration.

In addition, we consider that the density of energetic electrons is
much lower than that of the ambient electrons (i.e., $n_{be} \ll n_0$). Thus the dispersion
relation can be obtained approximately in terms of cold-plasma theory
\citep{mel86,che02p16,wud08p25}:
\begin{eqnarray}
N_ \pm ^2 = 1 - \frac{{\omega _{pe}^2}}{{{\omega _{k \pm }}({\omega
_{k \pm }} + {\tau _ \pm }{\Omega _e})}}
\end{eqnarray}
with ${\tau _ \pm } =  - {s_ \pm } \pm \sqrt {s_ \pm ^2 + {{\cos}^2}\theta}$ and ${s_ \pm } = {\omega _{k \pm }}{\Omega _e}{\sin^2}\theta /[2(\omega _{k \pm }^2 - \omega _{pe}^2)]$, where ${N_ \pm } = {k_ \pm }c/{\omega _{k \pm }}$ is the refractive index,  the subscript
``$+$" and ``$-$" denote O mode and  X (extraordinary) mode, respectively, $\theta$ is the propagation angle of the emitted electromagnetic waves with respect to the ambient magnetic field.

\section{Numerical results for solar coronal parameters}

\subsection{$\omega_{pe}/\Omega_{e}$ in the density-depleted duct}

From Equation (5), one can find that the density depletion is considerable when $v_{bi} \gg v_A$. However, quantitative description of the parameter $\omega_{pe}/\Omega_{e}$ depends on the ambient density as well as magnetic field in solar corona. For the sake of discussion, we describe the electron density profile by $n_0 = N_0 \times R^{-6}$ with its maximum $N_0$ having the typical value of $5 \times 10^9 ~ cm^{-3}$ in the corona base \citep{ver81p35,wuc02p94}, where $R$ is the heliocentric distance in units of solar radius ($R_{\sun}$). Such a density model is qualitatively similar to the Newkirk model \citep{new61p83} and compatible with the recent result of coronal density measurement for a active region \citep{cho13p48}, obtained from six filter images taken by the Atmospheric Imaging Assembly on board the Solar Dynamic Observatory \citep{lem12p17}. A model of unipolar spot-field configurations is chosen since coronal type II bursts are in general associated with active regions and strong magnetic field can be expected. That is \citep{gin64,yoo02p52}
\begin{eqnarray}
B=B_0(1-\frac{h}{\sqrt{h^2+b^2}}), ~~h=R-1,
\end{eqnarray}
where $B_0$, $h$ and $b$ are the maximum field intensity at the center of the spot on the photosphere, the height above the solar surface, and sunspot radius, respectively. Here both $h$ and $b$ are normalized by solar radius. It may be appropriate to let $B_0 = 1500$ G and $b=0.05$ for the specific choice \citep{fun66p59}. One can calculate the parameter $\omega_{pe}/\Omega_{e}$ in the duct via $\omega_{pe}=2\pi \times 8979 \sqrt{n_{D}}$ and $\Omega_{e}=2\pi \times 2.8 \times 10^6 B$, where $n_D$ is determined by Equation (5). Figure 3 presents the variation of the parameter $\omega_{pe}/\Omega_{e}$ with heliocentric distance in the duct. Here the assumption of $v_{bi}=10v_A$ has been used. It is clear that $\omega_{pe}/\Omega_{e} \lesssim 1$ can be satisfied at the distance below about 1.55 $R_{\sun}$.

\subsection{The growth rate of ECME}

Based on the Equations (8) and (9), one can study the ECME driven by beam electrons in the presence of AWs. It should be noted that the new resonant condition suggests that the number of harmonics is determined by the combinations of $n$ and $q$. The growth rate of the F will be calculated via setting $(n+q)=1$ which is the contributions of $(n,q)=(1,0)$ and $(0,1)$ while that of the H via setting $(n+q)=2$ which is the contributions of $(n,q)=(2,0)$, $(1,1)$ and $(0,2)$. In addition, the growth rate depends on two variables ($\omega_k$, $\theta$) for the fixed parameters. By the maximum growth rate, it means that the growth rate with the highest value in both ($\omega_k$, $\theta$). Finally, it is also considered that the emitting waves have the cutoff frequencies of $\omega_{oc} \simeq \omega_{pe}$ for the O mode and of $\omega_{xc} \simeq \sqrt{\omega_{pe}^2 + \Omega_e^2/4} + \Omega_e/2$ for the X mode.

\begin{figure}
\epsscale{1.0} \plotone{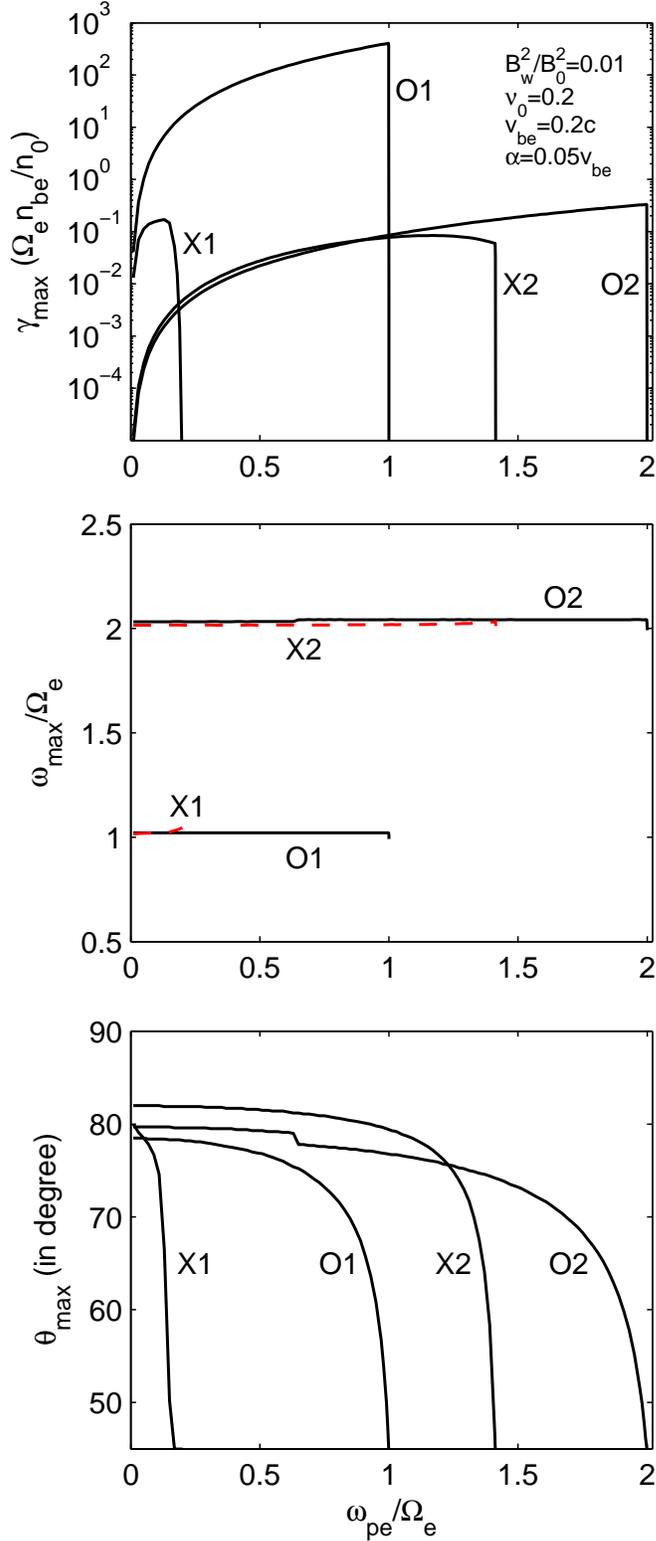} \caption{The $\gamma_{max}$, $\omega_{max}/\Omega_e$, $\theta_{max}$ vs.
$\omega_{pe}/\Omega_e$: O1, fundamental in the O mode;
O2, second harmonic in the O mode; X1, fundamental in the X mode; X2, second harmonic in the X mode.\label{fig4}}
\end{figure}

For the beam electrons accelerated by coronal shocks, they may have the typical energy of about 10 keV \citep[$ \sim 0.2c;$][]{dul08p43,man05p19,che11p01} or have higher energy \citep[up to $0.5c$;][]{cai87p65}. We will consider the beam has the velocity of $0.2c$ (i.e., $v_{be} = 0.2c$) in the calculation. For the sake of discussion, we assume that the AWs have the level of 0.01 relative to the ambient magnetic field (i.e., ${B_w^2}/{B_0^2}=0.01$). Figure 4 (top panel) plots the maximum growth rates calculated by varying the parameter $\omega_{pe}/\Omega_e$, where O1 and O2 are the F and H in the O mode, X1 and X2 the F and H in the X mode. The $\gamma_{max}$ is the maximum growth rate normalized by $\Omega_e n_b/n_0$. From Figure 4 (top panel), one can find that for the F the O1 mode is dominant when $\omega_{pe}/\Omega_e < 0.2$ and only the O1 mode is excited when $0.2 <\omega_{pe}/\Omega_e \lesssim 1$. For the H both the X2 and the O2 modes are excited in a larger range of parameter $\omega_{pe}/\Omega_e$ with the comparative growth rates. Here the parameter range of $\omega_{pe}/\Omega_e \lesssim 1$ is of interest in which both the F and H can be excited. The other two quantities ($\omega_{max}$, $\theta_{max}$) are the wave frequency corresponding to the maximum growth and the propagation angle at which the maximum growth occurs for a given parameter $\omega_{pe}/\Omega_e$. The middle and bottom panels display $\omega_{max}$ and $\theta_{max}$ versus $\omega_{pe}/\Omega_e$, respectively. A further study of wave propagation in the density-depleted duct is desirable and attention should be paid to this study in the future.

\section{Discussion related to solar type II radio bursts}

The study of solar type II radio bursts has a long history since their discovery \citep{pay47p56,wil50p87}. It is generally believed that these burst emissions are attributed to coronal shocks with a induced emission process \citep{wes57p87,wid59p76,uch60p76,cli99p89}. This idea is reasonable because (1) the propagation speed of the emission source is comparable to the shock speed; (2) the radiation usually occurs near solar active regions after a major flare or CME event in which a shock can be expected; (3) the induced emission, i.e., ECME in this paper, has a feature of narrow-band emission which coincides with the observations of type II bursts. Nevertheless, some basic facts of observations are still inexplicable.

An outstanding fact, similar to type III bursts \citep{dul80p03}, is that the positions of apparent sources of the F and H with the same frequency, as a general rule, are found to overlap \citep[see, a review by][]{nel85p33}. This result is inconsistent with conventional models implying that the source of the H should have a higher position than that of the F at a fixed frequency. Early some attempts, for the case of type III bursts, were made to explain this issue by refraction and by scattering off coronal irregularities \citep{ste71p62,rid72p48,rid72p98,leb73p41}. Later an alternate idea was proposed by \citet{dun79p89} in terms of wave ducting. The author assumed that radio emission generated at a lower height first propagates along under-dense magnetic flux tubes and then escapes from the under-dense flux tubes at a higher height. This process implies that the observed source position is not the position of wave generation, but the position of wave escape from the under-dense flux tubes. The key point is that the waves with the same frequency, either F or H, will escape nearly at the same position. It is thus conceivable that the apparent sources of the F and H are in general overlapping at a fixed frequency. In fact, the present model is inherently in line with this idea. As described in the Section 2, The ECME takes place in a density-depleted duct (responsible to under-dense magnetic flux tubes suggested by \citet{dun79p89}) in which the condition of $\omega_{pe}/\Omega_e \lesssim 1$ is fulfilled. The emitting waves have the frequencies near the local electron gyrofrequency and twice the electron gyrofrequency for the F and H respectively. Here it should be noted that the cutoff frequency of the exterior plasma, $\omega_{xc} \simeq \omega_{oc} \simeq \omega_{pe}$, is much higher than the frequencies of the emitted waves, since the plasma frequency of the exterior plasma is in general much higher than the local electron gyrofrequency for the corona of interest. The newly excited waves (either the F or H) hence cannot directly escape and will propagate inside the duct until they have arrived at certain heights where their frequencies slightly exceed the exterior cutoff frequency so that they can escape and become observable. This may be the reason that the observed sources of the F and H are usually almost coincident at the same frequency. In addition, it is clear that the present model is compatible with the observations of the height of type II bursts, suggesting that the radiation has a frequency close to the local plasma frequency, since the observed source is the exit point where the frequency of the escaping wave is nearly equal to the local ambient plasma frequency (i.e., cutoff frequency).

Another consequence of the present model is that the observed frequency drift of type II bursts reflects the decreasing magnetic field strength along the path of emission sources, rather than the density as assumed in conventional theories, since the frequency of an emitted wave is determined by the local electron gyrofrequency. Hence, the deduced velocity of emission sources will be determined by both the observed frequency drift rate and the assumed distribution of the ambient magnetic field along the path of emission sources. For a fixed drift rate, one can imagine that this velocity deduced from the magnetic field shall be smaller than that deduced from the density because the gradient of the magnetic field is, in general, larger than that of the density. For a case of type II burst reported by \citet{nak90p77}, an radial velocity of the emission source exceeding $10^4$ km/s was obtained by the authors according to the conventional density model \citep{new61p83}, while this velocity is about 3000 km/s calculated on the basis of the magnetic field model described by Equation (13).

\section{Discussion and conclusions}

This paper reveals the physical processes of radio emission from solar coronal shocks based on ECME, in which AWs are introduced and play an important role. The present discussion consists of three elementary processes based on (1) excitation of AWs, (2) depleting density and forming a duct, and (3) ECME in the duct. AWs is first excited by ion beams accelerated by NPSs. The generated AWs deplete the local density through magnetic compression in a low-beta plasma \citep{wuc06p17}. Consequently a density-depleted duct is made along the ion foreshock boundary where the ion beams have the highest energy \citep{eas05p41}. Then, the ECME works when energetic electrons with a crescent-shaped beam distribution move through the duct. As shown in the Section 3.2, the ECME is more efficient when the plasma frequency is smaller than the electron gyrofrequency (i.e., $\omega_{pe}/\Omega_e \lesssim 1$) in the duct. And, both the F and H are excited when $\omega_{pe}/\Omega_e \lesssim 1$ is fulfilled.

The present study is different from the preceding works \citep{wuc86p92,yoo07p01}. The key difference is that the present study introduces AWs which lead to three important plasma processes. They are the depleting density, pitch-angle scattering energetic electrons, and influencing the basic physics of ECME. Among these processes the first one is vitally significant because this process leads to the condition of $\omega_{pe}/\Omega_e \lesssim 1$ in favor of  the ECME to excite both the F and H. Furthermore, on the basis of density depletion a duct is made, which can inherently explain the outstanding fact that the observed source regions of the F and H of type II bursts are nearly overlapping at a fixed frequency \citep{saw82p49,nel85p33}. In addition, the present model implies that the observed radiation has a frequency close to the ambient plasma frequency of exit point, which is compatible with the observations of the height of type II bursts. Finally, it should be noted that the observed frequency drift of type II bursts is due to the decreasing magnetic field strength rather than the density as described in the section 4 based on the present model.

Certainly, the present discussion of radio emission from coronal shocks is preliminary. The radiation from shocks is by no means simple. Based on the preceding works \citep{wuc86p92,yoo07p01}, this paper mainly investigates the physical process of radio emission from solar coronal shocks and shows that the AWs play a vitally important role. More researches are required to fully understand the related details of the present discussion.

\acknowledgments
Research by D. J. Wu and G. Q. Zhao was supported by NSFC under grant Nos. 11373070 and 41074107, and by MSTC under grant No. 2011CB811402. Research by L. Chen was supported by NSFC under Grant Nos. 41304136, and by NSF of Jiangsu Province under Grant No. BK20131039. We are grateful to Dr. C. S. Wu and C. B. Wang for valuable discussions and comments on our work. We also thank the anonymous referee for helpful comments and suggestions.

\bibliography{myref}

\end{document}